\documentclass[aps,prl,reprint,groupedaddress,showpacs]{revtex4-1}
\usepackage{graphicx}
\usepackage{color}
\usepackage{amssymb}
\usepackage[tbtags]{amsmath}
\usepackage{bm}

\newcommand{\no}{\nonumber}
\newcommand{\ph}{^{\phantom{\dagger}}}

\begin{document}
\title{Strong Momentum-Dependent Electron-Magnon Renormalization of
a Surface Resonance on Iron}
\author{Beatrice Andres}
\affiliation{Freie Universit{\"a}t Berlin, Fachbereich Physik, Arnimallee 14,
14195 Berlin, Germany}
\author{Martin Weinelt}
\email[Corresponding author: ] {weinelt@physik.fu-berlin.de}
\affiliation{Freie Universit{\"a}t Berlin, Fachbereich Physik, Arnimallee 14,
14195 Berlin, Germany}
\author{Hubert Ebert}
\author{J{\"u}rgen Braun}
\affiliation{Ludwig-Maximilians-Universit{\"a}t M{\"u}nchen, Butenandtstr.\ 5-13, 81377 M{\"u}nchen, Germany}
\author{Alex Aperis}
\affiliation{Department of Physics and Astronomy, Uppsala University, P.\,O.\ Box 516, 75120 Uppsala, Sweden}
\author{Peter M. Oppeneer}
\affiliation{Department of Physics and Astronomy, Uppsala University, P.\,O.\ Box 516, 75120 Uppsala, Sweden}
\date{\today}

\begin{abstract}
\noindent
The coupling of fermionic quasiparticles to magnons is essential for a wide range of processes, from 
ultrafast magnetization dynamics in ferromagnets to Cooper pairing in superconductors. 
Although magnon energies are generally much larger than phonon energies, up to now their electronic band renormalization effect in ferromagnetic metals suggests a significantly weaker quasiparticle interaction. 
Here, using spin- and angle-resolved photoemission, we show an extraordinarily strong 
renormalization leading to replica-band formation of an iron surface resonance at $\sim$200\,meV. Its strong magnetic linear dichroism unveils the magnetic nature and momentum dependence of the energy renormalization. By determining the frequency- and momentum-dependent self-energy due to generic electron-boson interaction to compute the resultant electron spectral function, we show that the surface-state replica formation  is consistent with strong coupling to an optical spin wave in a Fe thin film.

\end{abstract}


\maketitle
The coupling between electrons and magnons is a fundamental quasiparticle interaction, which affects numerous ordering processes in magnetic 
materials. In spintronics, it plays an important role for spin-polarized tunneling \cite{Balashov_PRB78} and recent works give first evidence that magnons can drive ultrafast magnetization dynamics within 100 femtoseconds \cite{Haag_PRB90, Carpene_PRB91, Turgut2016, Eich2017, Tengdin2018}. In high-temperature superconductors spin fluctuations exist and paramagnons appear to be the most promising mediators for Cooper pairing in cuprates \cite{LeTacon2011}.  Consequently, the contribution of spin fluctuations to the electron self-energy in cuprates has been studied extensively \cite{Dahm2009,Kordyuk_PRL97,Borisenko_PRL96}. In contrast, only few works could capture the
influence of electron-magnon interaction in ferromagnetic metals \cite{Schaefer2004, Cui2007a, Mlynczak2019}. 

In photoelectron spectroscopy, electron-boson coupling mostly manifest itself as kinks in the electronic bandstructure. This quasiparticle renormalization is equally induced by interactions with phonons \cite{Rotenberg2000, Lanzara2001, Hofmann_BPW}, electrons \cite{Byczuk2007,Sanchez-Barriga_PRL103,Schaefer2005,Lonzarich_JMMM45}, magnons \cite{Schaefer2004, Cui2007a}, and plasmons \cite{Bostwick2007}. Apart from kinks in the electronic bands, strong electron-phonon coupling can also be reflected in replica band or satellite formation, which is rare, but was detected for high-temperature Fe-based superconductors \cite{Lee2014,Rebec2017,Choi2017} and TiO$_2$ \cite{Moser2013}.

The kinks in the renormalized bands allow us to quantify the electron-boson coupling strength. While in high-temperature superconductors the coupling strength $\lambda$ between electrons and spin fluctuations has been found to reach up to $\lambda = 0.95$ \cite{Kordyuk_PRL97}, the values reported for ferromagnets  are comparatively small.
Specifically, for Fe kinks were already found for surface \cite{Schaefer2004} and bulk iron states \cite{Cui2007a,Mlynczak2019}. The latter exhibit an energy renormalization at an energy of $\sim$270\,meV in the majority-spin channel near the Fermi energy \cite{Cui2007a}, and even at a binding energy of 1.5 eV in the minority-spin bulk state
\cite{Mlynczak2019}, while kinks in minority-surface resonances occur at a cutoff energy of $\sim$160\,meV \cite{Schaefer2004}. 
The coupling constants of the low-energy band renormalizations were estimated as
 $\lambda = 0.14\pm0.03$ \cite{Cui2007a} and $0.20\pm0.04$ \cite{Schaefer2004}, respectively. 
Comparing to manifestations of strong electron-phonon coupling, however, where besides kinks replica-band formation was observed \cite{Lee2014,Rebec2017}, such kind of distinctive band renormalization has not been reported for electron-magnon coupling.
  
\begin{figure}
	\includegraphics[width=\columnwidth] {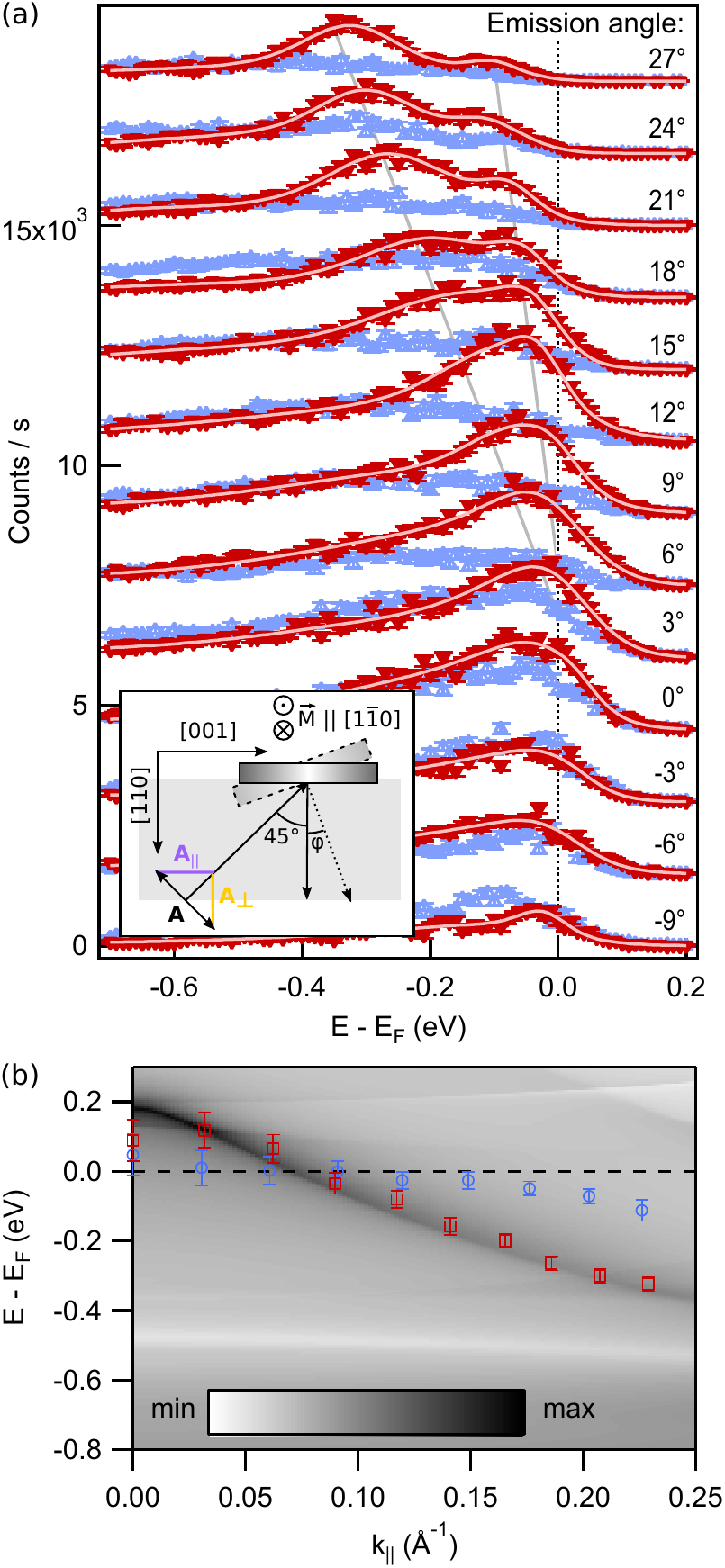}
	\caption{\label{fig:FesrLeft} Spin- and angle-resolved photoemission spectroscopy. (a) Energy distribution curves as a function of emission angle for minority- and majority-spin components (red tip-down and blue tip-up triangles).  The light gray lines are guides to the eye along the dispersion of the minority-spin peaks. The inset represents the measurement geometry in a schematic view on the light-incidence plane (gray). (b) Minority-spin photoemission intensities computed in a one-step photoemission calculation (dark gray band)  with the LSDA+DMFT approach. For comparison,
	the experimental peak positions are shown as red and blue symbols and have been extracted from fits to the minority-spin spectra (light red solid lines in (a)).}
	\end{figure}

In the present study, we use spin- and angle-resolved photoelectron spectroscopy to unravel the 
coupling between fermionic quasiparticles and magnons. For a 15 monolayer (ML) iron film on W(110), we study not only the dispersion and spin polarization of a renormalized surface resonance but also the magnetic linear dichroism (MLD) in spin-polarized photoemission. We observe a pronounced magnon-induced, momentum-dependent replica-band formation of the surface resonance.
MLD confirms the magnetic character of the band renormalization and confirms the momentum-dependence of the electron-magnon coupling. We show, by computing the quasiparticle dispersion due to different types of electron-boson interaction, that the observed band renormalization is consistent with strong electron-magnon coupling of the surface state to an optical spin wave in the thin Fe film.

In our spin-resolved photoemission experiment on a thin Fe film on W(110) 
we study the $\Sigma_{3,1}$ minority-spin surface resonance along the $\overline{\Gamma} -\overline{\mathrm{H}}$ direction of the surface Brillouin zone. We use linearly $p$-polarized light and a photon energy of 6.3\,eV with an exchange scattering spin detector \cite{Winkelmann2008}. The inset in Figure~\ref{fig:FesrLeft}(a) illustrates the experimental geometry. At normal electron emission ($\vartheta=0^\circ$), the light was incident at an angle of $45^\circ$, at which the vector field {\bf{A}} of the laser pulse contains equal parts of polarization parallel ($\bf{A}_{||}$) and perpendicular ($\bf{A}_{\perp}$) to the sample surface. Rotation of the sample to larger emission angles $\vartheta$ increases $\bf{A}_{||}$ and reduces $\bf{A}_{\perp}$. According to dipole selection rules, with a $\Sigma_1$-symmetric final state, electrons can be photoemitted from $\Sigma_1$ states by the $\bf{A}_\perp$ component and from $\Sigma_3$ states by the $\bf{A}_{||}$ component \cite{Hermanson1977}. A series of two-photon-photoemission measurements confirming the $\Sigma_{3,1}$ spatial symmetry character of the observed surface resonance is presented in the  Supplemental Material (SM), Fig.~S3.

The dispersion of the surface resonance along $\overline{\Gamma} - \overline{\mathrm{H}}$ is shown in Fig.~\ref{fig:FesrLeft}(a), in a series of spin-resolved spectra measured at different polar angles $\vartheta$ from $-3^\circ$ to $27^\circ$. To analyze the origin of the minority-spin band splitting, we perform calculations based on the local spin-density approximation plus dynamical mean-field theory (LSDA+DMFT) approach (see SM). A one-step photoemission calculation  of the minority-spin intensity for a photon energy  of 6.3\,eV is shown in Fig.~\ref{fig:FesrLeft}(b), which corresponds well with the measured main peak positions of the surface resonance (red squares). 
The surface resonance disperses downward in energy and thereby parallels the dispersion of the related $\Sigma_{3,1}$ bulk band. The strong minority-spin character of the  surface resonance (see SM, Fig.~S1 for extracted spin polarizations) corroborates the spin-resolved photoemission and inverse photoemission results in Refs.~\cite{Sanchez-Barriga_PRL103} and \cite{Braun_PRB65}, respectively. The one-step photoemission calculations in Ref.~\cite{Braun_PRB65} furthermore explain how the corresponding \textit{majority-spin} surface resonance is shifted above the Fermi level $E_F$ and is not detected in our experiment. 

\begin{figure}[tb!]
	\includegraphics[width=1.0\columnwidth]{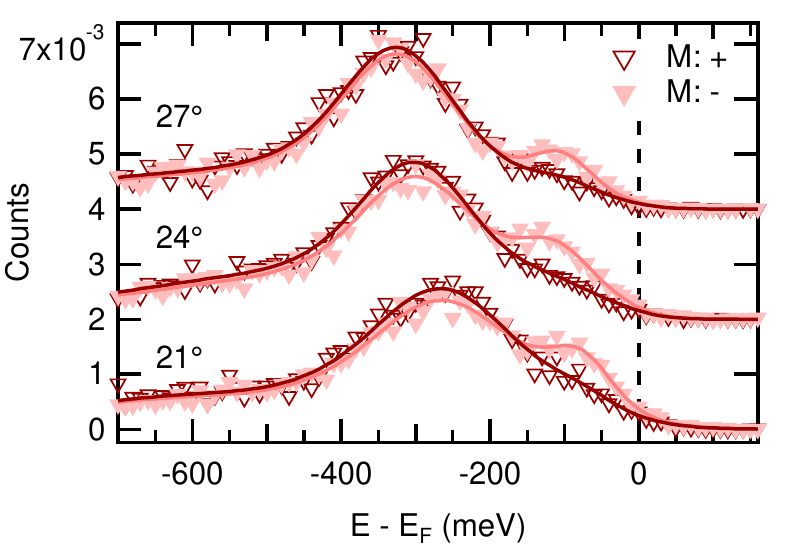}
	\caption{\label{fig:Dichroism} Pairs of minority-spin spectra for reversed sample magnetization directions M+ and M-, measured at emission angles of 21$^\circ$, 24$^\circ$, and 27$^\circ$. Total peak areas have been normalized for equal spectral weight. The reversal of the magnetization direction changes the spectral weight of both peaks leading to the magnetic linear dichroism in the minority spin channel.}
\end{figure}

In photoemission the surface resonance appears at 0.08\,\AA$^{-1}$ ($\sim$9$^\circ$ emission angle, Fig.~\ref{fig:FesrLeft}(a)). For emission angles $>12^\circ$, the minority-spin surface resonance clearly splits into two peaks, one with a steeper and one with a flatter dispersion meeting at $E_F$ (indicated by the gray lines in Fig.~\ref{fig:FesrLeft} (a)). The theoretical photoemission spectra in Fig.~\ref{fig:FesrLeft}(b), derived from one-step photoemission calculations and based on the LSDA+DMFT calculated band structure and wave functions, predict that there is only one single dispersing band and no second band to explain the flatter dispersing feature in the measured spectra (blue circles). Furthermore, symmetry considerations allow us to exclude that the two dispersing branches stem from different states. Knowing that both states have either $\Sigma_3$ or $\Sigma_1$ spatial symmetry and are of minority-spin character, the corresponding bands should avoid a crossing and form a hybridization gap \cite{Rampe_PRB57}. Such gaps have been found to be on the order of $\approx 100$\,meV \cite{Pickel2008}. From the fact that, in our spectra, there is no gap but rather an enhanced intensity at the crossing point, we can exclude the two peaks to stem from different electronic states. 
Symmetry considerations hence corroborate that both peaks must originate from the same surface resonance, \textit{i.e.}, the single undressed surface band as given by the LSDA+DMFT calculations develops  through quasiparticle renormalization  into two dispersing branches.

Next, we discuss the occurrence of MLD appearing jointly with the energy renormalization. The measurement geometry in Fig.~\ref{fig:FesrLeft}(a) fulfills the general requirement $|\textbf{A} \cdot (\textbf{M} \times \textbf{k})| > 0$ for MLD in 
photoemission, which is similar to the transversal magneto-optical Kerr effect. 
Fig.~\ref{fig:Dichroism} shows pairs of \textit{minority-spin} spectra with the sample magnetized in opposite directions, \textit{i.e.}, for spin magnetic moments aligned antiparallel to  M+ and M-, respectively. 
These pairs of minority spin spectra have each been measured at a fixed angle in the $k_{||}$ range, in which we find both peaks simultaneously. The intensity of the  renormalized state closer to $E_{\rm F}$ increases when the magnetization is reversed, the intensity of the main peak decreases slightly. A complete  $E(k_{||})$-map of MLD in the minority-spin channel can be found in the SM, Fig.~S4a.

Since MLD is a signature of coupling between spin and momentum space, the observed MLD 
is as well a consequence of the electron-magnon coupling. Regarding the spin system, spin-polarized electron energy loss spectroscopy (SPEELS) \cite{Zakeri_PRL104,Zakeri_PhysRep545}, gives already evidence for an asymmetric coupling of magnons to the electronic system. This shows up in different magnonic lifetimes for opposite wave vectors \cite{Zakeri2012}, which have been assigned to the asymmetric Dzyaloshinskii-Moriya exchange interaction -- just like the Rashba splitting observed in the magnon bandstructure \cite{Zakeri_PRL104}. Regarding the electronic system, this asymmetry will influence the magnon-induced renormalization in the photoemission spectrum, leads to the observed redistribution of the spectral weight between dressed and undressed states, and corroborates the magnetic nature of the electron dressing.


The LSDA+DMFT calculations account for electron-electron interaction and give the bare minority-spin surface state, but do not take into account quasiparticle renormalization due to electron-phonon, magnon, or plasmon interactions. To analyze the effect of electron-boson renormalization we start with considering a generic interaction Hamiltonian,
\begin{eqnarray}\label{h0}
{\cal H} &=&\sum_{{\bf k}}\varepsilon_{\bf k}^{ }c^\dagger_{{\bf k}}c\ph_{{\bf k}} + \sum_{{\bf q},\nu}\hbar\omega_{{\bf q}\nu}\Big(b^\dagger_{{\bf q}\nu}b\ph_{{\bf q}\nu}+\frac{1}{2}\Big) 
\nonumber \\
& &+
 \sum_{{\bf q},\nu}\sum_{{\bf k}}I\ph_{\nu}c^\dagger_{{\bf k+q}}c\ph_{{\bf k}}\left(b\ph_{{\bf q}\nu}+b^\dagger_{{\bf -q}\nu} \right)\, ,
\end{eqnarray}
where the first (second) term describes the electron (boson) kinetic energy with band energy $\varepsilon_{\bf k}$ and boson energy $\hbar \omega_{{\bf q}\nu}$, respectively. The third term is the potential energy due to their mutual interaction with interaction strength $I_{\nu}$. Using the Matsubara formalism on the imaginary axis and analytic continuation to the real frequency axis, 
the electron-boson interaction leads to the 
 real-frequency-dependent electron self-energy at temperature $T$ \cite{Marsiglio1988,Marsiglio1990},
\begin{eqnarray}\label{selfenergy}
\Sigma({\bf k},\omega) \!=\!-T \!\!\sum_{n=-\infty}^{\infty}\sum_{{\bf k'},\nu}I_\nu^2\Big\{D_\nu({\bf q},\omega-i\omega_{n}) G({\bf k'},i\omega_n)
-
\nonumber \\
\!\!\!  \sum_\pm \pm \frac{G_R({\bf k'},\omega\pm\omega_{{\bf q}\nu})}{2}\Big(\tanh{\frac{\omega\pm\omega_{{\bf q}\nu}}{2T}}\mp\coth{\frac{\omega_{{\bf q}\nu}}{2T}}\Big)\Big\},~~~
\end{eqnarray}
where
\begin{eqnarray}\label{propagator}
D_\nu({\bf q},i\omega_n-i\omega_{n'})=\frac{-2\omega_{{\bf q}\nu}}{(\omega_n-\omega_{n'})^2+\omega^2_{{\bf q}\nu}}
\end{eqnarray}
is the branch-resolved magnon Matsubara propagator for fermionic Matsubara frequencies,
$ \omega_n = (2n +1) \pi k_B T $. $G({\bf k},i\omega_n)$, $G_R({\bf k},\omega)$ are the Matsubara frequency dependent  and retarded, real-frequency dependent full Green functions. The latter is given by
\begin{eqnarray}\no
G_R^{-1}({\bf k},\omega)&=&[G^R_0]^{-1}({\bf k},\omega)-\Sigma({\bf k},\omega)\,,
\end{eqnarray}
with the bare Green function,
$G^R_0({\bf k},\omega)=[\omega+i\delta-\varepsilon_{\bf k}]^{-1}$,
where $\delta$ is small and corresponds to physical broadening.

\begin{figure}[tb!]
	\includegraphics[width=0.85\linewidth] {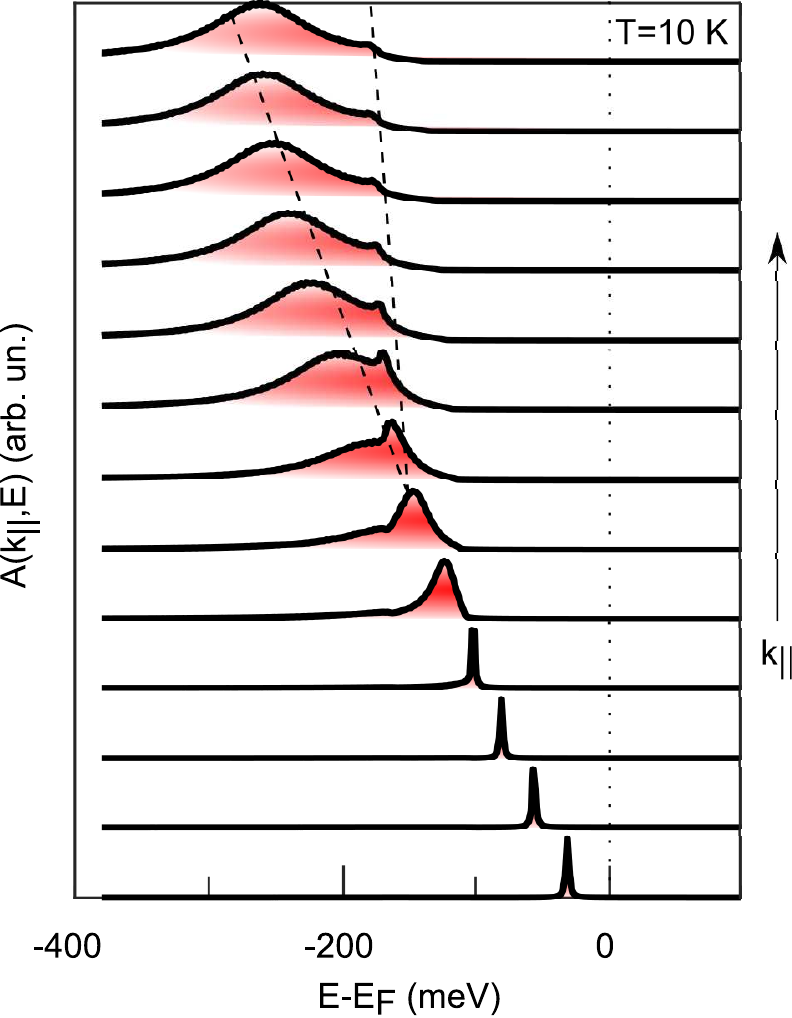}
	\caption{\label{fig3} 
	Computed quasiparticle spectral density due to electron-magnon coupling to an optical  magnon branch that has a characteristic frequency in the range $120 - 160$ meV.  The interaction strength $I_o$ is here 110\,meV.}
\end{figure}

We solve Eq.\ (\ref{selfenergy}) with a one-step calculation, employing the one-loop approximation:
\begin{eqnarray}\label{oneloop}
G_R^{-1}(\Sigma)
\approx G^{-1}_0-\Sigma(G_0)\,,
\end{eqnarray}
which is reasonable since we are interested in qualitative features of the spectra due to poles of the electronic spectral function.
Subsequently, from Eq.\ (\ref{oneloop}) we calculate the electron spectral function,
$A({\bf k},\omega)=- {\pi}^{-1}{\rm Im} \,G_R({\bf k},\omega)$,
which is proportional to the photoemission spectra. 
%
We approximate the bare electron dispersion in Fig.\ \ref{fig:FesrLeft}(b) as
\begin{eqnarray}\label{xk}
\varepsilon_{\bf k}=t\cos ({\bf k}a_0)-\mu ,
\end{eqnarray}
with $t=200$\,meV, $\mu=300$\,meV and consider next its coupling to specific magnon branches.
The spin wave spectra of ultrathin Co and Fe films were extensively investigated in the past
\cite{Zhang2009,Zakeri_JElSpecRelPh189,Zakeri_PhysRep545,Chen2017}. Due to the spatial confinement thin Fe bcc(110) films exhibit acoustic as well as gapped optical spin wave branches, that can be parametrized using Heisenberg exchange constants,
\begin{eqnarray}\label{magnons}
\! \omega_a({{\bf q}})\!=\!12 {J_N}\big[1-\cos{(\frac{{{\bf q}}a_0}{2})}\big]\!+ \!4 {J_{NN}} \!\left[1-\cos{({{\bf q}}a_0)}\right] ,~~\\
\! \! \omega_o({\bf q})\!=\! 4{J_N}\big[3-\cos{(\frac{{\bf q}a_0}{2})}\big] \!+ \!4{J_{NN}} \!\left[3-\cos{({\bf q}a_0)}\right], ~~
\end{eqnarray}
where ${{\bf q}}={\bf q}_\parallel$, the in-plane momentum, $a_0=3.165$\,\AA\  is the lattice constant, and
 ${J_N}=7.6$\,meV and ${J_{NN}}=4.6$ meV are the nearest and next-nearest neighbor exchange constants \cite{Zhang2009} (see SM, Fig.~S5 for a sketch of the dispersion relations). 
Next, we compute the spectral function considering acoustic or optical magnons, or a dispersionless Einstein phonon mode with frequency $\Omega_0=40$\,meV, while varying the interaction strength $I_{\nu}$ and using $\delta = 1.5$\,meV. The spectral function computed for an optical magnon branch, shown in Fig.\ \ref{fig3},  
provides consistent agreement with experiment. Results for band renormalization due to the acoustic magnon and Einstein phonon modes are given in the SM, Fig.~S6. The interaction with the optical magnon mode causes an unusual, strongly momentum-dependent renormalization of the surface state. While in the past a kink has been detected \cite{Schaefer2004}, here we find a \textit{splitting} of the bare electron dispersion in two quasiparticle bands. Specifically, the spectral density at lower binding energy is a momentum-dependent replica of the state at higher binding energy, a feature that has not been reported before for electron-magnon coupling. It is similar to the appearance of replica-band formation \cite{Moser2013,Lee2014,Rebec2017,Choi2017} due to electron-phonon interaction \cite{Aperis2018,Schrodi2018}, where, however, it is practically momentum independent.

Previously, the bulk electron-magnon coupling constant $\lambda$ of Fe was estimated from the  ARPES-measured self-energy $\Sigma ({\bf k}, \omega )$, giving $\lambda \approx 0.14$ \cite{Cui2007a}. For the surface resonance it was estimated from the kink in the band dispersion
\cite{Schaefer2004}. However, this procedure is less accurate and cannot be employed in our case. We directly compute the coupling constant $\lambda_\nu$ by the general formula
\begin{eqnarray}
\label{lambda}
\lambda_\nu=N_0\Big\langle \Big \langle\frac{2I^2_\nu}{\omega_{{\bf q}\nu}}\Big \rangle_{{\bf k}^{\prime}_{F}}\Big \rangle_{{\bf k}^{ }_{F}}
\end{eqnarray}
with $N_0=0.0012$ states/meV the density of states at the Fermi level of the electron band  and $\langle\ldots\rangle_{{\bf k}_{F}}$ 
a momentum average over the Fermi surface. This procedure yields 
$\lambda_o=0.2$ for the optical magnon.
Although at first sight this value does not seem large compared to strong electron-phonon coupling constants $\lambda \approx 1$, it deserves to be noted that the magnon frequency in the denominator of Eq.\ (\ref{lambda}) is much larger than typical phonon frequencies. Recalculating the coupling constant for a 40 meV-phonon frequency gives that it would correspond to an electron-phonon coupling with $\lambda = 0.7$, thus indicating a comparably strong coupling. 
Lastly, as was shown previously, these spin waves are strongly damped \cite{Costa2003,Zhang2012}. The robust coupling of electrons to these optical magnons could provide a relevant channel for ultrafast laser-induced demagnetization \cite{Beaurepaire1996,Bea_PRL115,Tengdin2018} in thin films that has so far not been considered.

In summary, we find a remarkable quasiparticle renormalization of the minority-spin surface resonance on Fe(110). 
The renormalization occurs in the typical energy range ($160 - 200$ meV) for magnonic excitations in thin Fe/W(110) films \cite{Chuang_PRL109} and its magnetic nature is further corroborated by a redistribution in the MLD spectral weight upon reversal of the sample magnetization.
Our findings particularly reveal that not only in high-temperature Fe-based superconductors but also in an elementary ferromagnet electronic states can show exceptional
coupling to spin excitations. This adds new knowledge regarding the coupling to magnons and spin fluctuations and contributes to obtaining a better understanding of ultrafast spin dynamics in thin magnetic systems.

\begin{acknowledgments}
B.A., M.W.\ and P.M.O.\ acknowledge the Collaborative Research
Center TRR 227 ``Ultrafast Spin Dynamics" for financial support. J.B.\ and H.E.\ acknowledge longstanding support from the Deutsche Forschungsgemeinschaft (SPP1666 priority program (Grants No.\ EB 154/32 and EB 154/37)), and the Bundesministerium f\"ur Bildung und Forschung through
BMBF: 05K16WMA for financial support. A.A.\ and P.M.O.\ acknowledge support from the Swedish Research Council (VR), the K.\ and A.\ Wallenberg Foundation (grant No.\ 2015.0060), the R{\"o}ntgen-{\AA}ngstr{\"o}m Cluster, and the Swedish National Infrastructure for Computing (SNIC).
\end{acknowledgments}

%

\end{document}